\documentclass[a4paper, 11pt]{article}

\usepackage[utf8]{inputenc}
\usepackage[T1]{fontenc}

\usepackage[ttscale=.875]{libertine}
\usepackage[libertine]{newtxmath}

\usepackage[margin=2.5cm]{geometry}
\usepackage{booktabs,dcolumn,rotating}
\usepackage{microtype}
\usepackage{color}
\usepackage{amsfonts}
\usepackage{graphics}

\begin{document}

\title{\textbf{Insufficient properties of image encryption algorithms}}

\author{Martin Stanek \\[1ex]
  {\small Department of Computer Science} \\ 
  {\small Comenius University} \\
  {\small\mbox{stanek@dcs.fmph.uniba.sk}}}
\date{}
\maketitle

\begin{abstract}
We analyze the security of recently proposed image encryption scheme \cite{ECCHC}. 
We show that the scheme is insecure and the methods used to evaluate its security 
are insufficient. By designing the Deliberately Weak Cipher, a completely vulnerable 
cipher with good statistical properties, we illustrate our main point -- a solid 
analysis cannot be replaced by some selected set of statistical properties.
\end{abstract}


\section{Introduction}

Encryption is an important tool for ensuring confidentiality of data. Recently,
the ECCHC -- a combination of elliptic-curve cryptography and Hill cipher --
was proposed by Dawahdeh, Yaakob and bin Othman \cite{ECCHC}. 
The proposal is specifically aimed at image encryption.
The authors justify the security of the ECCHC by evaluating selected statistical
properties on sample plaintext and ciphertext images. The properties are entropy, 
peak signal to noise ratio, and unified average changing intensity.

\paragraph{Our conribution.} We analyze the security of the ECCHC scheme and show
its various weaknesses that render the scheme unusable for any security sensitive
application. Evaluating some selected set of statistical 
properties is not an adequate replacement for analyzing how cryptanalytic 
attacks apply to an encryption scheme. In order to accentuate this point, we propose a
toy example -- the Deliberately Weak Cipher (DWC). The DWC is completely weak
(with only $8$ bit key and other serious vulnerabilities), but it attains
comparable or even better statistical parameters (depending on input image type) 
than the ECCHC.

\section{The ECCHC scheme}

The ECCHC scheme combines public key cryptography based on elliptic curves with symmetric 
encryption in a straightforward way. The authors tailored their proposal to encryption
of $8$-bit grayscale images of size $256\times 256$ pixels, although it can be easily
extended to other image types. To keep our presentation simple, we also use this 
particular image type. The weaknesses identified in the scheme are relevant for 
possible extension as well.

\subsection{Overview of the scheme}

The ECC part of the scheme is basically a Diffie-Hellman key agreement. Let 
$(E,\cdot)$ be a group of points on some elliptic curve, generated by a generator 
$G$. Let $|E| = p$ be a prime. A user $U$ can generate his/her private key $n_U$ by 
randomly choosing from $\{0,1,\dots,p-1\}$. The corresponding public key is 
$P_U =n_U G$.

Two users, $A$ and $B$, can agree on a shared point $K_I = (x,y) \in E$ by computing
$K_I=n_AP_B = n_An_BG$ (for user $A$) or $K_I=n_BP_A = n_An_BG$ (for user $B$). The 
point $K_I$ is used to obtain a $2\times 2$ matrix $K$:

$$
  K=\begin{bmatrix} k_{11} & k_{12} \\ k_{21} & k_{22} \end{bmatrix},
$$
where $(k_{11},k_{12}) = xG$ and $(k_{21},k_{22}) = yG$. A $4\times 4$ self-invertible
matrix $K_m$ is formed as follows ($I$ denotes a $2\times 2$ identity matrix):

$$
  K_m=\begin{bmatrix} K & I-K \\ I+K & -K \end{bmatrix},
$$
Since the matrix is used for encrypting plaintexts with the alphabet of size $256$ 
($8$-bit grayscale) the elements are transformed mod $256$.

The second part of the scheme is Hill cipher used in the ECB (Electronic Codebook) 
mode to sequentially encrypt $4\times 1$ vectors from a plaintext image. 
We denote by $P_1,\ldots, P_n$  the sequence of $4\times 1$ vectors that 
the image is split into (for $256\times 256$ image we have $n=256^2/4=16384$). 
The ciphertext is a sequence of blocks 
$C_1,\ldots,C_n$, where the $i$-th block/vector $C_i$ is computed by simple 
matrix-vector multiplication: $C_i = K_m\cdot P_i$. Given that $K_m$ is 
self-invertible, a decryption uses the same multiplication: $P_i = K_m\cdot C_i$.

\subsection{Weaknesses of the ECCHC scheme}
\label{sec_HCprop}

We discuss some properties of the ECCHC that, in our opinion, make the scheme insecure
and unsuitable for any security sensitive application.

\paragraph{Low entropy of the symmetric key.} The matrix $K_m$ is uniquely determined
by any of its quadrants, e.g. $4$ elements of $K \bmod 256$ are sufficient to 
reconstruct $K_m$. Hence the worst case complexity of brute-force attack is very
low $\sim 2^{4\cdot 8}= 2^{32}$.

\paragraph{Known plaintext attack.} The proposal inherits the linearity of Hill cipher.
This makes it vulnerable to known plaintext attack. For $4\times 1$ plaintext
block and corresponding ciphertext block we obtain four linear equations with 
$4$ unknowns (values of the matrix $K$). For random plaintext block we can find 
a unique solution with high probability. Multiple distinct plaintext-ciphertext pairs
of blocks yield the key with certainty.

\paragraph{Using the symmetric transformation in the ECB mode.} Opting for the ECB mode
has a well known weakness: equal plaintext blocks are encrypted to equal ciphertext
blocks, $P_i=P_j \; \Rightarrow \; C_i=C_j$. This property can be particularly unfortunate
for images containing patterns or drawings, where equal $4\times 1$ blocks are expected.
It is certainly possible to modify the symmetric part of the ECCHC to employ a counter
or using other modes, however the proposal does not address this issue in any way.

\paragraph{Fixed points.} The structure of self-invertible matrix $K_m$ guarantees that
vectors of the form $P=(p,p,p,p)^T$, for any $p$, are fixed points regardless of
actual values in $K_m$:

$$
 K_m\cdot P = \begin{bmatrix} K & I-K \\ I+K & -K \end{bmatrix} \cdot 
              \begin{bmatrix} p \\ p \\ p\\ p \end{bmatrix}
            = \begin{bmatrix} K \cdot \begin{bmatrix} p\\p \end{bmatrix} + 
                              (I-K) \cdot \begin{bmatrix} p\\p \end{bmatrix} \\ 
                              (I+K) \cdot \begin{bmatrix} p\\p \end{bmatrix}
                              -K \cdot \begin{bmatrix} p\\p \end{bmatrix} \end{bmatrix}
            = \begin{bmatrix} p \\ p \\ p \\ p \end{bmatrix}.
$$

\begin{figure}[h]
\centering\includegraphics[width=90px]{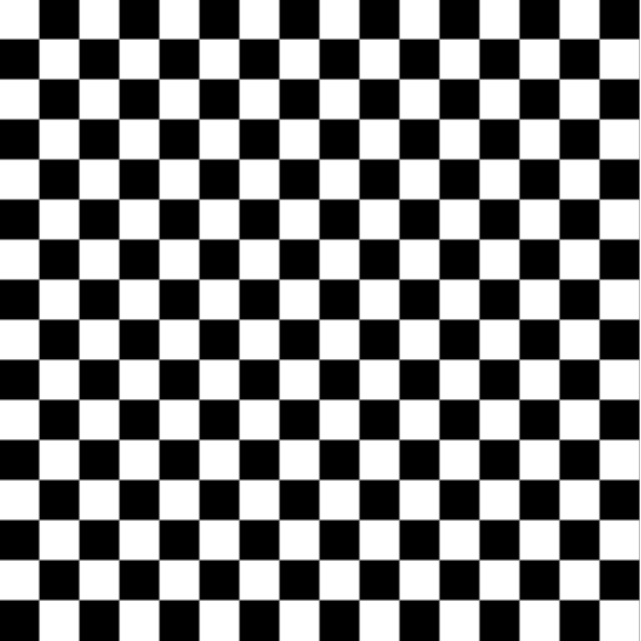}
\caption{Checkerboard -- invariant image for any symmetric key $K_m$.}
\label{fig1}
\end{figure}

\begin{figure}[h]
	\centering\includegraphics[width=90px]{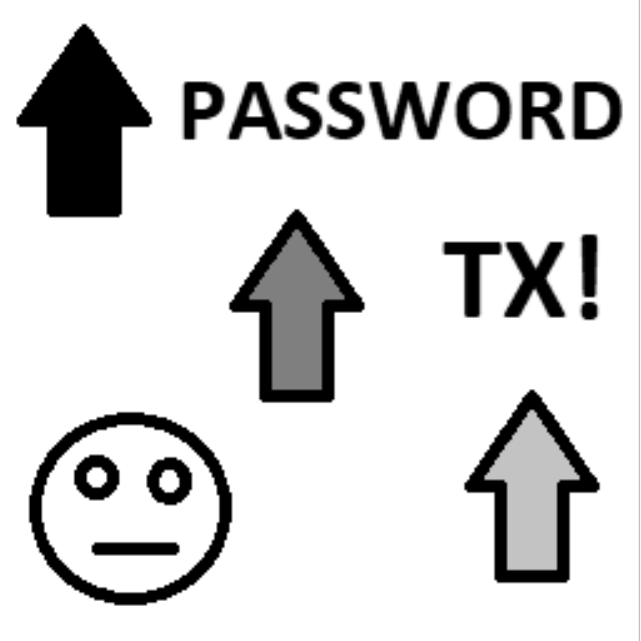}\hskip6em
	\includegraphics[width=90px]{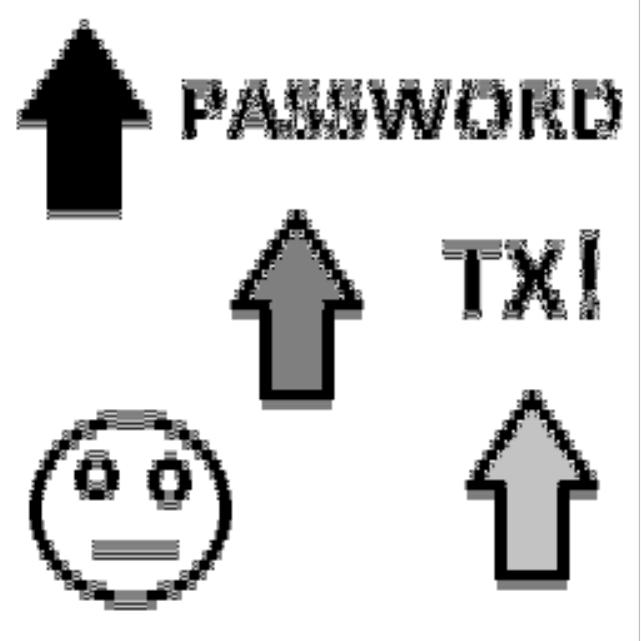}
	\caption{ECCHC -- Plaintext and ciphertext of a drawing.}
	\label{fig2}
\end{figure}

Let us illustrate the ECB mode and the fixed points issues. A checkerboard image, see Figure \ref{fig1},
is an example of plaintext image that remains intact after encrypting with any $K_m$. 
Encrypted drawing with uniform color areas might be comprehensible (only slightly distorted 
along edges). An example is shown in Figure \ref{fig2}.

\paragraph{Inadequate security analysis.} The security analysis included
in the proposal contains evaluation of these three statistical properties on few
sample images:

\begin{itemize}
\item Entropy of the encrypted image considering frequencies of pixel values.
\item Peak Signal to Noise Ratio (PSNR) using original and encrypted images.
\item Unified Average Changing Intensity (UACI) using original and encrypted images.
\end{itemize}

\noindent Such analysis neglects other cryptographic properties the symmetric cipher is expected
to satisfy. More importantly, it ignores all potential attacks (some weaknesses were discussed
in previous paragraphs). To further illustrate this point we propose an intentionally weak
cipher (weaker than the symmetric part of the ECCHC) and show that we can obtain comparable
or even better values of above properties, see Section \ref{sec-DWC}.

\paragraph{Unclear public key part of the ECCHC.} The elliptic curve key agreement part
of the ECCHC is under-specified. The proposal does not define what kind of elliptic curves
should be used, how large the parameters should be (small entropy of symmetric key $K_m$ indicates
that using standardized elliptic curves is pointless), how exactly are $xG$ and $yG$ 
values transformed into matrix $K$ (when to apply mod $256$ operation), etc.

\section{Deliberately weak cipher}
\label{sec-DWC}

We designed the Deliberately Weak Cipher (DWC for short) to illustrate the inadequacy
of statistical measures for assessing the strength of encryption algorithms. 
Suitable values of statistical properties are necessary but by no means sufficient
condition for secure encryption. We start by reviewing properties used in the ECCHC
proposal.

\subsection{Remarks on statistical properties}
\label{sec-statprop}

Entropy for $8$-bit grayscale image $X$ is calculated as $\text{Entropy}(X) = -\sum_{i=0}^{255} 
p_i \log_2 p_i$, where $p_i$ denotes a fraction of pixels with value $i$. The maximum entropy 
in this case is $8$, and we expect that random noise image has entropy close to $8$.

The Peak Signal to Noise Ratio (PSNR) is used to measure quality between signal with a noise 
and the original signal. In our case, let $A = (a_{i,j})_{i,j=1}^{n}$ and $B=(b_{i,j})_{i,j=1}^{n}$
be two $8$-bit grayscale images. The PSNR, expressed in decibels, is computed as follows:
\[
  \text{PSNR}(A,B) = 20\cdot \log_{10} \frac{255}{\sqrt{\text{MSE}(A,B)}}\, ,
\]
where $\text{MSE}(A,B)=n^{-2}\cdot \sum_{i,j=1}^{n} (a_{i,j}-b_{i,j})^2$ is the Mean Square Error 
between these two images. It is easy to calculate the expected value of the PSNR between all-black 
(or all-white) image and a random noise image -- the MSE is $21717.5$, 
and the corresponding PSNR is $\sim 4.7627$. Similarly, the expected value of the PSNR between two 
random noise images is $\sim 7.7476$.

The Unified Average Changing Intensity (UACI) measures the average distance among pixel values
using following formula:
\[
  \text{UACI}(A,B) = \frac{1}{n^2}\cdot \sum_{i,j=1}^{n} \frac{|a_{i,j}-b_{i,j}|}{255}\;\times 100\%.
\]
Trivially, the UACI of all-black (or all-white) image and a random noise image is $50$\%. 
Calculating the expected value of the UACI for two random noise images yields $1/3 + 1/(3\cdot 255)
\sim 33.46$\%.

We are interested in two cases: (1) a pair of two random noise images, and (2) a pair of
monochrome black (white) image and a random noise image. The distinction between these cases 
is important in the evaluation of encryption algorithms. Consider a cryptographically 
strong encryption algorithm and pair of plaintext and ciphertext images. We expect that 
a statistical properties for this pair are close to the first case if the plaintext is 
a photograph, painting, etc.; and close to the second case if the plaintext is a drawing,
checkerboard, or simply an image with a vast majority of black/white pixels.

\subsection{The DWC algorithm}

The DWC encryption uses $8$-bit key $k$, obviously too short for any serious application.
To make it similar to the ECCHC, a core transformation used in the DWC takes $4\times 1$ input 
vector. Let $P=(p_0, p_1, p_2, p_3)^T$ be an input vector of four bytes. The core transformation
is defined as follows:
\[
  \text{CT}(P) = \mathbf{M} \odot (S(p_0), S(p_1), p_2, S(p_3))^T,
\]
where $S$ is an $8$-bit s-box and $\mathbf{M}\, \odot$ computes a fixed matrix 
multiplication, borrowed from the AES SubBytes and MixColumns transformations,
respectively \cite{FIPS-197}. Please note that 
our matrix multiplication operates on a single vector (column), instead of four 
columns transformation MixColumns in the AES. 
The output of CT is again $4\times 1$ vector. 
The core transformation can be easily inverted by multiplying with inverse matrix 
and performing three lookups to the table representing $S^{-1}$. Although we do not 
care about the performance of the DWC, various enhancements in this area can be 
obtained from vast literature on the implementation of the AES.

To mimic the ECCHC we split an image into a sequence of $4\times 1$ vectors 
$P_1,\ldots, P_n$. The ciphertext is a sequence of blocks $C_1,\ldots,C_n$:
\[
  C_i = \text{CT}(P_i \oplus i \oplus ((k \oplus \text{lsb}(i)) \ll 24)), \qquad \text{for } i=1,\ldots,n,
\]
where $\oplus$ denotes bitwise XOR operation, lsb returns the least significant byte, and
$x \ll 24$ shifts the byte $x$ to the left for $24$ position. Thus $(k \oplus \text{lsb}(i))
\ll 24)$ yields a 32-bit vector with the most significant byte being $k \oplus \text{lsb}(i)$
and other three bytes being $0$. 
The decryption is straightforward: $P_i = \text{CT}^{-1}(C_i) \oplus i \oplus 
((k \oplus \text{lsb}(i)) \ll 24))$, for $i=1,\ldots,n$.

\subsection{Properties of the DWC algorithm}

The DWC algorithm is very weak. The most important vulnerability is small key space 
(overall, only $256$ keys), and thus being susceptible to brute-force attack. 
Another weaknesses follow from fact that the core transformation CT is fixed and 
does not depend on the key in any way. Therefore anyone can compute 
$\text{CT}^{-1}$ on ciphertext blocks and obtain 
$P_i \oplus i \oplus ((k \oplus \text{lsb}(i)) \ll 24))$. Since $i$ is known (it is 
just a counter) one can easily compute $P_i \oplus (k \ll 24))$. This is the original
plaintext block with the most significant byte XOR-ed with $k$. Hence anyone can 
recover $75$\% of the image without knowing the key.

\begin{figure}[h]
	\centering
	\includegraphics[width=90px]{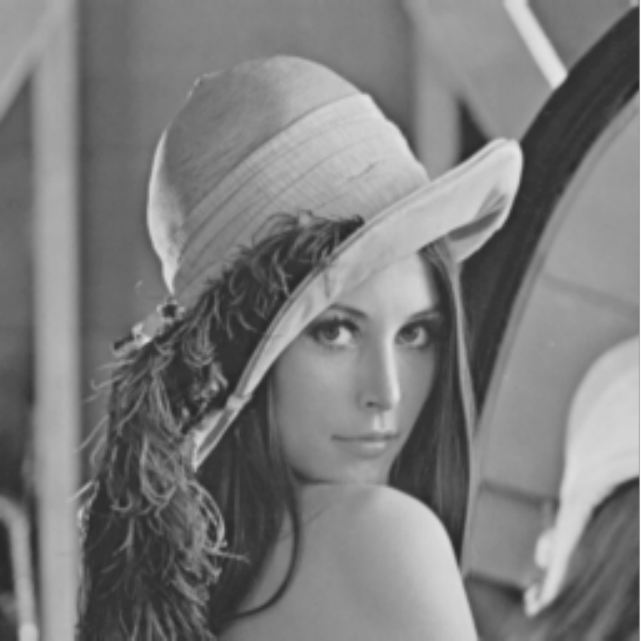}\hskip6em
	\includegraphics[width=90px]{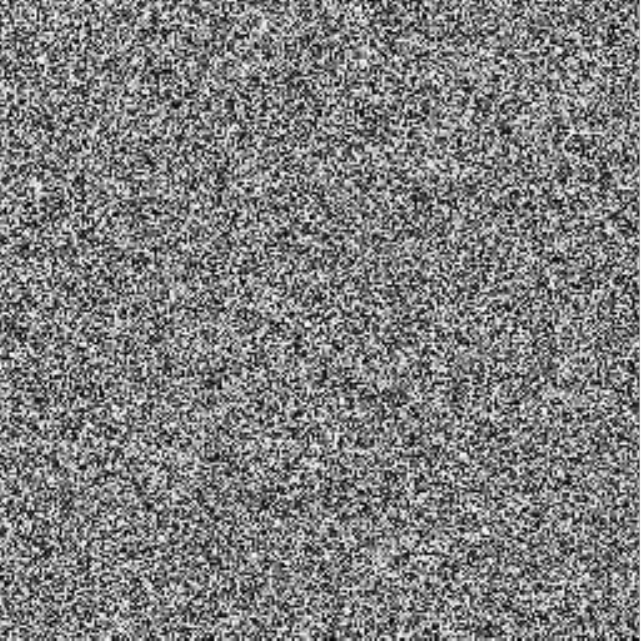} \\[1em]
	\includegraphics[width=90px]{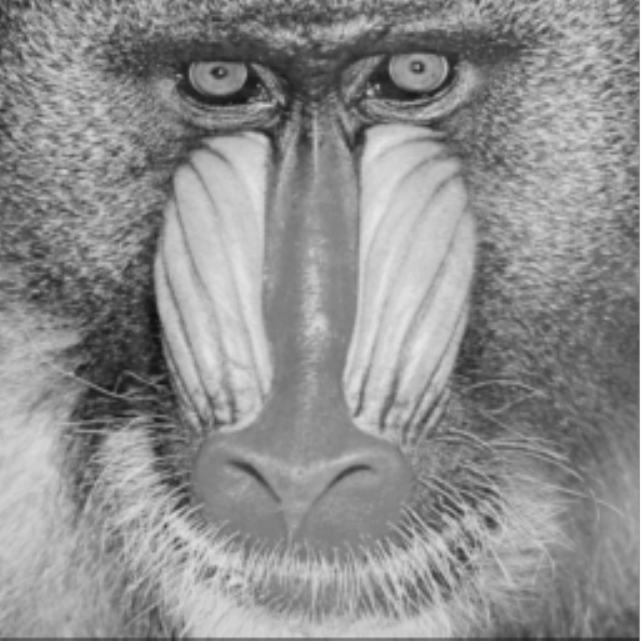}\hskip6em
	\includegraphics[width=90px]{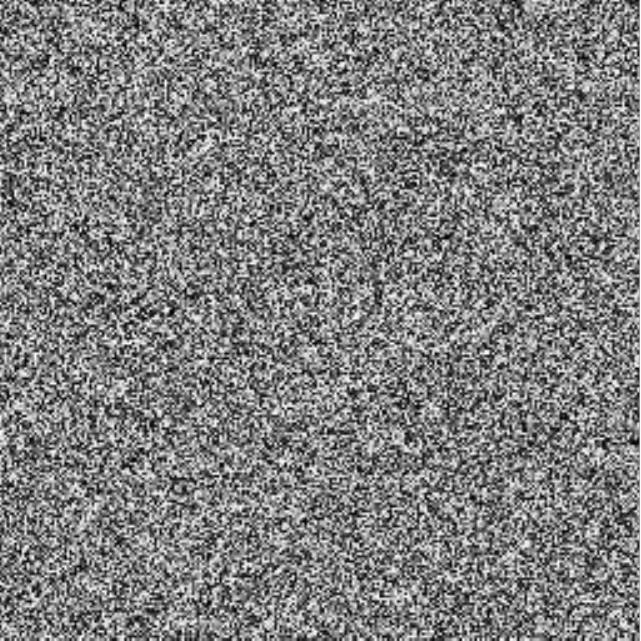} \\[1em]
	\includegraphics[width=90px]{checkerboard.pdf}\hskip6em
	\includegraphics[width=90px]{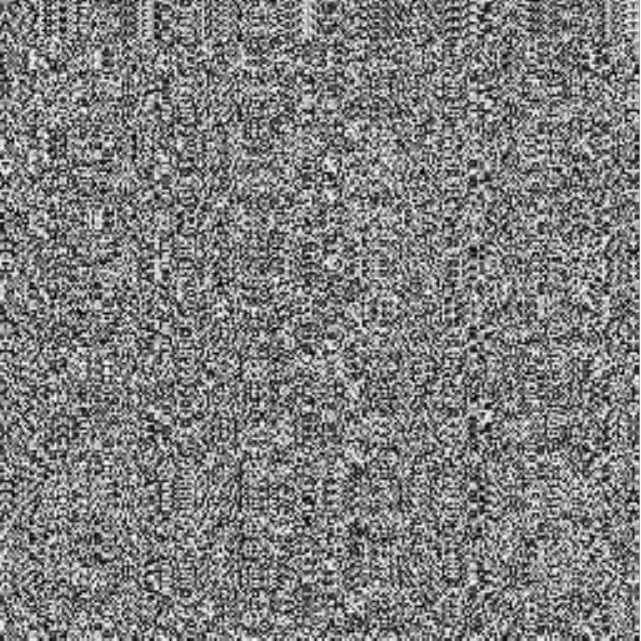} \\[1em]
	\includegraphics[width=90px]{draw-plain.pdf}\hskip6em
	\includegraphics[width=90px]{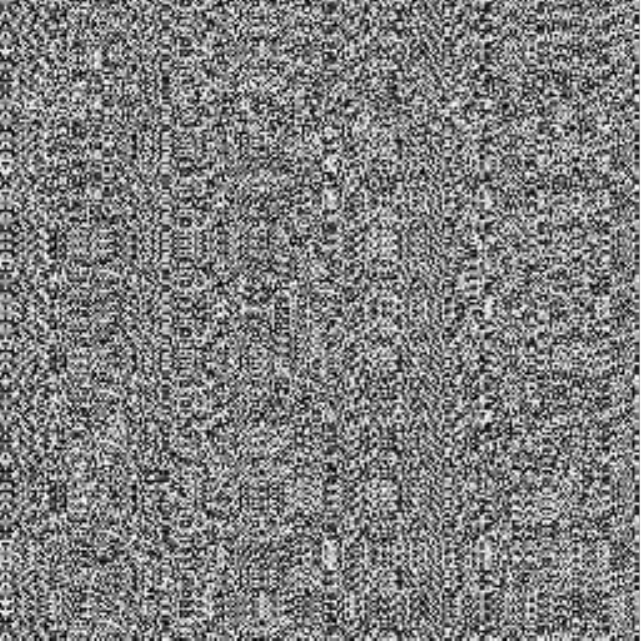} 
	\caption{DWC on sample images.}
	\label{fig3}
\end{figure}

Design of the DWC is aimed at obtaining good values of statistical properties, such as the entropy,
the PSNR and the UACI. We use four sample images -- two standard photographs (lena, baboon), 
checkerboard, and drawing (both used in Section \ref{sec_HCprop} as well). 
Figure \ref{fig3} shows a visual overview of plaintext and corresponding ciphertext 
images. There are no fragments of original images, visible patterns or obvious 
regularities in the ciphertext images.

Table \ref{tab-compare} compares the numerical values of the entropy, the PSNR and the UACI
for the DWC and the ECCHC for our sample images. The results for lena and baboon images
are very close, and we can consider the DWC and the ECCHC as equal in these types of images.
Image types like checkerboard or drawings pose a problem for the ECCHC (as already 
discussed in Section \ref{sec_HCprop}). On the other hand, the DWC produces 
ciphertext images with much superior statistics -- close to values we expect 
for those image types, see Section \ref{sec-statprop}. 

We can summarize -- the DWC is very weak cipher that excels in the entropy, the PSNR,
and the UACI properties.

\begin{table}[!ht]
\centering
\begin{tabular}{llrrr}
  \toprule
  algorihtm & image & Entropy & PSNR & UACI [\%] \\
  \midrule 
  ECCHC 
  & lena 			& $7.9947$ & $9.3709$ & $28.1994$ \\
  & baboon 			& $7.9965$ & $9.4589$ & $28.0132$ \\
  & checkerboard	& $1.0000$ & $\infty$ & $0.0000$ \\
  & drawing  		& $1.6929$ & $16.2244$ & $3.9406$ \\
  \midrule 
  DWC
  & lena 			& $7.9974$ & $9.4180$ & $28.1236$ \\
  & baboon 			& $7.9971$ & $9.5001$ & $27.8896$ \\
  & checkerboard	& $7.9979$ & $4.7623$ & $50.0049$ \\
  & drawing  		& $7.9979$ & $4.9161$ & $48.8540$ \\
  \bottomrule 
\end{tabular} 
\caption{Comparison of ECCHC and DWC for sample images}
\label{tab-compare}
\end{table}

\noindent\emph{Remarks.} (1) The key length of the DWC could be even smaller. 
We can fix $k$ to some constant and get an encoding scheme good statistical properties
and no security at all. \\ 
(2) The DWC is by no means a unique construction. Various approaches can be used to
fulfill the same goals. For example, many lightweight stream ciphers with 
severely reduced key length would be similarly vulnerable, while still having
good values of the entropy, the PSNR and the UACI for image encryption.

\section{Conclusion}

We showed various weaknesses of the ECCHC image encryption scheme, despite
good statistical properties published in the proposal \cite{ECCHC}. 
More importantly, we illustrated insufficiency of such approach to analyzing
the strength of encryption algorithm by proposing the Deliberately Weak Cipher.
The DWC has comparable or even better statistical parameters (depending on 
input image type) than the ECCHC, while being completely weak.

For any encryption scheme proposal, the real cryptanalytic assessment 
of the scheme should be conducted. Evaluating some selected set of statistical 
properties is not an adequate replacement for analyzing how cryptanalytic 
attacks apply to the scheme.

The ECCHC should not be used for image (and other data) encryption. Use 
standard encryption techniques. In case of resource-constrained devices,
the area of lightweight cryptography offers alternative algorithms with better
security. For recent survey on lightweight cryptography, see \cite{LC17}.



\begin{thebibliography}{99}

\bibitem{ECCHC}
Dawahdeh Z.E., Yaakob S.N., bin Othman R.R.,
\emph{A new image encryption technique combining Elliptic Curve Cryptosystem with 
Hill Cipher},
Journal of King Saud University -- Computer and Information Sciences,
Volume 30, Issue 3, 2018, pp. 349-355.
\mbox{{https://doi.org/10.1016/j.jksuci.2017.06.004}}

\bibitem{LC17}
McKay K.A, Bassham L., Turan M.S., Mouha N.,
\emph{Report on Lightweight Cryptography}, NISTIR 8114, 2017.
\mbox{{https://doi.org/10.6028/NIST.IR.8114}}

\bibitem{FIPS-197}
National Institute of Standards and Technology (NIST),
\emph{Advanced Encryption Standard (AES)}, Federal Information Processing Standard
(FIPS PUB) \#197, 2001.
 
\end{thebibliography}
\end{document}